\documentclass[10pt]{article}  
\newcommand{\comma}{\, , \; \; }
\newcommand{\period}{\, .}
\newcommand{\eq}{\; = \;}
\newcommand{\sep}{\, , \;\;}

\newcommand{\be}{\begin{equation}}
\newcommand{\bd}{\begin{displaymath}}
\newcommand{\ee}{\end{equation}}
\newcommand{\ed}{\end{displaymath}}
\newcommand{\ba}{\begin{eqnarray}}
\newcommand{\ea}{\end{eqnarray}}

\newcommand{\minus}{\! - \!}

\renewcommand{\i}{{\rm i}}

\newcommand{\infsm}{{\scriptstyle \infty }}
\newcommand{\e}{{\rm e}}

\newcounter{storeeqn}
\renewcommand{\theequation}{\arabic{section}.\arabic{equation}}

\title{Spontaneous magnetization of the \\superintegrable 
chiral Potts model:\\ calculation of the determinant
$D_{PQ}$}

\author{ R.J. Baxter\\
{\protect \small  Mathematical
Sciences Institute,  The Australian National}\\
{\protect  \small  University,
 Canberra, A.C.T. 0200, Australia, \small e-mail: none }}

\date{\protect \small  21 December 2009
\\
revised 11 January 2010}

\begin{document}


\maketitle

 \abstract{For the Ising model, the calculation of the spontaneous 
magnetization leads to the problem of evaluating a determinant. Yang 
did this by calculating the eigenvalues in the large-lattice limit.
Montroll, Potts and Ward expressed it as a Toeplitz determinant and 
used Szeg{\H o}'s theorem: this is almost certainly the 
route originally  travelled by 
Onsager. For the corresponding problem in the superintegrable chiral 
Potts model, neither approach appears to work: here we show
that the determinant $D_{PQ}$ can be expressed as that of a product of 
two Cauchy-like matrices. One can then use the elementary exact 
formula for the Cauchy  determinant. One of course regains the known 
result, originally conjectured in 1989. }


 \vspace{5mm}




 \section{Introduction}
\setcounter{equation}{0}
Extrapolating from the Ising case and from series expansions, 
Albertini {\it et al} conjectured in 1989 \cite{AMPT89} that the 
order parameter or spontaneous magnetization 
of the solvable $N$-state chiral Potts model is
\be  \label{mag}
{\cal M}_r \eq (1-k'^2)^{r(N-r)/2 N^2} \period \ee
Here $r$ is an integer, between 0 and $N$, and $k'$ is a parameter 
that is ``universal'' in the sense that it is the same for all 
rows and columns of the lattice, even for an inhomogeneous
model where the rapidities vary from row to row and from 
column to column.\cite{BPAuY88} Here we consider the 
ferromagnetically ordered phase of the system,
where $0 < k' < 1$. This $k'$ is small at low temperatures
(high order), and tends to one at the critical temperature
(vanishing order), so we can regard it as a ``temperature''.

The author was able to derive  the result (\ref{mag})  in 
2005,\cite{RJB2005a,RJB2005b} using an analytic method based on 
functional relations satisfied by generalized order parameters
in the large-lattice limit.

If the vertical rapidities of the homogeneous model take a 
particular value, or if the those of a model with alternating 
vertical rapidities satisfy a certain 
relation,\cite{Baxter1988, Baxter1989} then we obtain the 
``superintegrable'' case of the chiral Potts model.
For this case, Gehlen and 
Rittenberg~\cite{vonGehlen1985}
showed that the horizontal and vertical components of the 
transfer matrix satisfy the ``Dolan-Grady'' 
condition,\cite{DolanGrady1982}
which  ensures that they generate the Onsager 
algebra. This is the algebra generated by the transfer matrices
of the Ising model.\cite[eqs. 59-61]{Onsager1944},
\cite[eq. 4.12]{AuYangPerk1989},
\cite{Davies1990} - \cite{AuYangPerk2010b}
Onsager used it to calculate the free energy of the Ising model.

In fact the superintegrable chiral Potts model looks very much like 
the Ising model. (For $N=2$ it {\em is} the Ising model.)
If one considers the model on a cylinder of
$L$ columns, 
with the spins on the top and bottom rows fixed to the value zero, 
then the row-to-row transfer matrices can be reduced from dimension
$N^L$ to dimension $2^m$, where $m$ is not greater than $L$. Like 
the Ising  model, the partition function is a matrix element of 
a direct product of $m$ matrices, each of dimension two.

For the Ising model, one can calculate the correlations
quite explicity as determinants by using free-fermion 
operators,\cite{Kaufman1949,Yang1952}
or equivalently by writing the partition functions directly as 
pfaffians.\cite{MPW1963}

If the partition functions of the $N$-state superintegrable model 
resemble those of the Ising model, then perhaps the
correlations are also similar, and can be obtained
by similar methods.
In particular, the  order parameter can be defined as 
\be  \label{prob}
{\cal M}_r \eq  \langle \omega^{r a} \rangle \comma \ee
where $r =1, 2, \ldots, N-1$, 
\be \omega \eq \e^{2 \pi \i /N} \comma \ee
and $a$ is a particular spin  inside the lattice. We can use this 
definition for a finite lattice: we do so herein. 
We only expect the simple result
(\ref{mag}) to be true in the limit when the lattice is large
and the spin $a$ is deep inside it. It should then be independent
of the values of the rapidity parameters, so should be 
the same for the general solvable model as for the 
superintegrable case.

We take the spins to have the values $ 0, \ldots, N-1$
and $\sigma = \{ \sigma_1, \ldots ,\sigma_L \}$ to be 
the set of spins in a horizontal row of the lattice. Let
$u_b$ be the $N^L$-dimensional vector with entries
\be (u_b)_\sigma  \eq \delta(\sigma_1,b) \, \delta(\sigma_2,b)
\cdots \delta(\sigma_L ,b)   \ee
and $S_r$ the diagonal matrix
with entries
\be \label{defS}
(S_r )_{\sigma, \sigma'}  \eq \omega^{r \sigma_1}
\prod_{j=1}^L \delta (\sigma_j, \sigma'_j) \period \ee
Then with these boundary conditions, 
(\ref{prob}) can be written in terms of the row-to-row transfer 
matrix $T$ as
\be  \label{partfn1}
 {\cal M}_r \eq \frac{u_0^{\dagger}\,  T^m S_r
 T^n \, u_0}{ u_0^{\dagger} \, T^{m+n} \, u_0}
\comma \ee
where $m$ ($n$) is the number of rows below (above) the 
particular spin $a$. We have chosen $a$ to lie in the first 
column: since we are using cylindrical (cyclic) boundary
conditions, this is no restriction.

The tranfer matrix $T$ commutes with a hamiltonian $\cal H$.
For convenience, we replace (\ref{partfn1}) with
\be  \label{partfn2}
 {\cal M}_r^{(1)} \eq \frac{u_0^{\dagger}\,  
\e^{-\alpha {\cal H}} S_r  \, \e^{-\beta {\cal H}} \, u_0}
{ u_0^{\dagger} \, \e^{-(\alpha+\beta) {\cal H}} \, u_0}
\period \ee
In the limit of $m, n$ large, the only eigenvectors of $T$
entering the RHS of (\ref{partfn1}) are those 
corresponding to the $N$ asymptotically degenerate largest
eigenvalues. The same is true of (\ref{partfn2}) in the limit
of $\alpha, \beta$ large and positive. Hence these limits
of (\ref{partfn1}), (\ref{partfn2}) must be the same.

Let $R$ be the operator that increases every spin in a row by one:
its elements are
\be R_{\sigma, \sigma'} \eq \prod_{j=1}^L \delta 
(\sigma_j, \sigma'_j +1 )  \period \ee
Since $R^N=1$, its eigenvalues are $1, \omega, \omega^2, \ldots,
\omega^{N-1}$. Let ${\cal V}_P$ (for $P = 0, 1, \ldots, N-1$) 
be the space of vectors $v$
such that  
\be  R v \eq \omega^P v \period \ee Then the full 
$N^L$-dimensional space is the union of ${\cal V}_0, \ldots ,
{\cal V}_{N-1}$. Let $v_P$ be the vector
\be v_P \eq N^{-1/2} \sum_{b=0}^{N-1} \omega^{-Pb} \, u_b 
\period \ee
Then $v_P$ and $\e^{-\beta {\cal H}} \, v_P$ are in ${\cal V}_P$.
If \be \label{QPr}
Q = P+r \; \;  , \; {\rm mod}\; N   \comma \ee
 then
${\cal S}_r \, \e^{-\beta {\cal H}} \, v_Q$ is also in 
${\cal V}_P$. Vectors in different spaces
 ${\cal V}_P$,  ${\cal V}_Q$ are orthogonal.
It follows that, for $b = 0, \ldots N-1$,
 \addtocounter{equation}{1}
 \setcounter{storeeqn}{\value{equation}}
 \setcounter{equation}{0}
 \renewcommand{\theequation}
{\thesection.\arabic{storeeqn}\alph{equation}}
 \be \label{uu}
u_b^{\dagger}\,  
\e^{-\alpha {\cal H}} S_r  \, \e^{-\beta {\cal H}} \, u_0 \eq 
N^{-1} \sum_{P=0}^{N-1} \omega^{-P b} \, W_{P,P+r} 
(\alpha, \beta ) \ee
\be u_b^{\dagger} \, \e^{-\alpha {\cal H}} \, u_0 
\eq N^{-1} \sum_{P=0}^{N-1} \omega^{-P b} \,  Z_P (\alpha) \comma \ee
\setcounter{equation}{\value{storeeqn}}
 \renewcommand{\theequation}{\arabic{equation}}
 \noindent where
\addtocounter{equation}{1}
\setcounter{storeeqn}{\value{equation}}
 \setcounter{equation}{0}
\renewcommand{\theequation}
  {\thesection.\arabic{storeeqn}\alph{equation}}
\be \label{defWPQ}
W_{PQ} (\alpha, \beta ) \eq v_P^{\dagger} \, 
\e^{-\alpha {\cal H}} S_r  \, \e^{-\beta {\cal H}} \,v_Q 
\comma \ee
\be \label{defZP}
Z_P(\alpha) \eq v_P^{\dagger} \, \e^{-\alpha {\cal H}} \,  
v_P \period \ee
\setcounter{equation}{\value{storeeqn}}
 \renewcommand{\theequation}{\thesection.\arabic{equation}}

Each LHS of (1.12) is the partition function of a lattice
where the top spins are fixed to the value 0, the bottom to the value
$b$. If $b \neq 0$ this ensures that there is at least one
 mis-matched seam (between phases where most spins are zero and 
most spins have value $b$)
running horizontally across the lattice. If 
$\zeta$ is the interfacial tension, (which we expect to be  
independent of $L$) then this will make
each partition function smaller then that for $b=0$ by  
a factor $\e^{-L \zeta}$.\cite[section 7.10]{book} 


In the limit of $L$ large the ratio of 
these expressions for $b \neq 0$
to their values for $b=0$ will therefore become zero. From (1.12),
 it follows that in this limit $Z_P (\alpha)$ is independent of
$P$, while $W_{PQ} (\alpha, \beta ) $ depends on $P, Q$ only 
via their difference $r = Q-P$. We show at the end of section 5 that 
these assertions are certainly true in the limit when 
$\alpha, \beta, L$ all tend to infinity.

The numerator in (\ref{partfn2}) can therefore be replaced by
$W_{PQ}(\alpha, \beta)$, for any $P,Q$ satisfying
(\ref{QPr}). The denominator can be replaced by
$Z_P(\alpha+\beta)$, for any $P$, but if $\alpha$ is also large,  
each $Z_P (\alpha)$ 
is of the form $K \e^{\nu \alpha}$, where $K, \nu$ must be independent
of $P$, so we can more symmetrically replace the denominator by
 $[ Z_P(2 \alpha) Z_Q (2 \beta)]^{1/2}$, giving
as our final expression for the spontaneous magnetization
\be  \label{partfn3}
 {\cal M}_r^{(2)} \eq \frac{W_{PQ}(\alpha, \beta) }
{ [ Z_P(2 \alpha) Z_Q (2 \beta)]^{1/2}} 
\period \ee
The three expressions $ {\cal M}_r, \,  {\cal M}_r^{(1)}, 
 \, {\cal M}_r^{(2)}$
are equal in the limit $L, m, n$ or $L, \alpha, \beta$
all becoming infinite.



\subsection*{Previous calculations for finite $L, \alpha, \beta$}


In \cite{paper1,paper2,paper3}, 
 we have looked at the problem of calculating 
$W_{PQ}(\alpha, \beta), Z_P(\alpha)$, and
hence ${\cal M}_r^{(1)}$, 
${\cal M}_r^{(2)}$,  algebraically,  for the superintegrable 
chiral Potts model, with {\em finite} $L, \alpha, \beta$.
We refer to these papers as I, II, III, and prefix their equations
accordingly.

The calculation of $Z_P(\alpha)$ is straightforward, being a minor
adaptation of the partition function calculations of 
\cite{Baxter1988, Baxter1989}, and is given in paper II.
The real problem is to calculate $W_{PQ}(\alpha, \beta)$,
or equivalently the ratio
\be \label{Dratio}
{\cal D}_{PQ}(\alpha, \beta) \eq \frac{W_{PQ}(\alpha, \beta)}
{Z_P (\alpha) \, Z_Q (\beta) } \period \ee
If we also define
\be  \label{defztilde}
{\cal  Z}_P(\alpha ) \eq Z_P(2 \alpha)/Z_P(\alpha)^2 \comma \ee
then (\ref{partfn3}) becomes 
\be \label{partfn4}
{\cal M}_r^{(2)} \eq \frac{{\cal D}_{PQ}(\alpha, \beta) }
{{[\cal  Z}_P(\alpha ) {\cal  Z}_Q(\beta )  ]^{1/2}} \period \ee


In I we considered the $N= 2$ case, which is the Ising model.
We used Kaufman's spinor operators (Clifford algebra) \cite{Kaufman49}
to first write ${\cal D}_{PQ}(\alpha, \beta)$ (for $P=0, Q=r=1$) 
as the square root of an  $L$ by $L$ determinant in I.4.59. 
Then in section 6 of I, eqn. I.6.29 and I.7.9, 
we further reduced this result to an $m$ by $m$ 
determinant (with no square root), where 
$m \leq L/2$. With obvious modifications of notation 
to allow for the working of the later papers,
and taking $\rho=0$, $x=1$ in I.3.5, I.3.6 
(also in II.5.23 and II.5.25),  this result can 
be written as
\be \label{defD}
{\cal D}_{PQ}(\alpha, \beta) \eq 
\det [ I_m - X_P(\alpha) E_{PQ} B_{PQ} X_Q(\beta)
E_{QP} B_{QP} ] \comma \ee
where $I_m$ is the identity matrix of dimension $m$,
$X_P(\alpha), E_{PQ}, X_Q(\beta), E_{QP}$ are diagonal matrices, 
$B_{PQ}$ is an $m$ by $m'$ matrix, where $|m-m'| = 0$ or 1,
$B_{QP}= -B_{PQ}^T$ 
and
$B_{PQ}$ is orthogonal in the sense that
\be \label{orth}
B_{PQ}^T B_{PQ} = I_{m'} \; \; {\rm if} \; m  \geq m' 
\sep
B_{PQ} B_{PQ}^T = I_{m} \; \; {\rm if} \; m  \leq m' \period \ee
We used the first ($L$ by $L$) form to take the
limit $\alpha,\beta \rightarrow + \infty$: the result of course
agreed with that of Yang\cite{Yang1952} and Montroll, Potts and 
Ward\cite{MPW1963}, and with (\ref{mag}) above.

In II we considered the 
superintegrable chiral Potts model and showed that the 
$N^L$-dimensional
matrices in (\ref{defWPQ}) could be replaced by ones
of lower dimension. In particular, the ${\cal H}$ in the first 
exponential could be reduced to dimension $2^m$,
where (for $P = 0, \ldots,  N \! - \! 1$),
 \be m = m_P = \; \; {\rm integer \; part \; of \;}
\left[ \frac{(N-1)L-P}{N}\right] \period \ee
Similarly, the second $\cal H$ could be replaced by
one of dimension $2^{m'}$, where
$m' = m_Q$, and ${\cal S}_r$ by a
$2^m$ by $2^{m'}$ matrix $S_{PQ}$. 
(The $p,q,  {\cal S}^r_{\rm red} $ of paper II became 
$P,Q, S_{PQ}$ in paper III and herein.)


We went on to  conjecture that (\ref{defD}), (\ref{orth})
also applied  to the superintegrable model, with fairly
obvious generalizations of the definitions of the
$X, E, B$ matrices.
We observed 
that this conjecture agreed with numerical tests performed 
to 60 digits  of accuracy.

These calculations involved sets of $m$ quantities
$\theta_1, \ldots ,\theta_m$ defined by
\be \label{defcos}
\cos \theta_j \eq c_j \eq (1+w_j)/(1-w_j) \comma
\; 0 \leq \theta_i < \pi \comma \ee
where $w_1, \ldots , w_m$ are the zeros of the
$m$-th degree polynomial  
$\rho_{\raisebox{-2pt}{$\scriptstyle  P$}} (w)$ 
given by
\be \label{defrho}
\rho_{\raisebox{-2pt}{$\scriptstyle  P$}}(z^N) \eq z^{-P} 
\sum_{n=0}^{N-1} \omega^{n P}
\, \left( \frac{1-z^N}{1- \omega^{-n} z} \right) ^L \comma \ee
taking $w = z^N$. Let  $c = (1+w)/(1-w)$ and, for all 
complex numbers $c$,
\be \label{defpoly}
{\cal P}_P (c) \eq N^{-L} (c+1)^m \; 
\rho_{\raisebox{-2pt}{$\scriptstyle  P$}}(w)   \period \ee
Then ${\cal P}_P (c) $ is the polynomial with zeros 
$c_1, \ldots ,c_m$, i.e.
\be \label{prodc}
{\cal P}_P (c)  \eq \prod_{j=1}^m (c-c_j) \period \ee
Similarly, we can define $m'$ quantities
$\theta'_1, \ldots ,\theta'_{m'}$ and 
$c'_1, \ldots ,c'_{m'}$ by replacing $P$ by $Q$ in the above 
three equations.
The $\rho_{\raisebox{-2pt}{$\scriptstyle  P$}}(w)$, 
$ {\cal P}_P (c)$ here are those of paper III, 
which are the $P(w)$,
$ {\tilde P}_p (c)$ of II.2.17, II.2.18 and  II.6.4.

In III we showed that both the ${\cal S}_r$ and 
$S_{PQ} $   matrices satisfied various
commutation relations with the hamiltonians, in particular that
$S_{PQ}$ satisfied III.3.39 and III.3.40.
We conjectured in III.3.45 that the elements of $S_{PQ}$
were simple ratios of products of trigonometric functions of
the $\theta_i$. We suggested 
that this result applied for any values of the $\theta_i$ 
and $\theta'_i$, not
necessarily those given by (\ref{defcos}) and (\ref{defrho}).
 Further,
if we defined ${\cal D}_{PQ}$ by III.3.48, then it 
was also given as a determinant by III.4.9 and III.4.10,
again for arbitrary  $\theta_i$,  $\theta'_i$.


\subsection*{Outstanding problems and progress}


It therefore appears that there is indeed an algebraic
route to calculating ${\cal D}_{PQ}$. However, there are 
still three outstanding problems to be overcome:

1.  To prove that the elements of $S_{PQ}$ are given by
III.3.45, and hence  ${\cal D}_{PQ}$ by III.3.48.

2. To further prove that ${\cal D}_{PQ}$ is given as a 
determinant by III.4.9 or equivalently III.4.10.

3. To calculate the determinant III.4.9 in the limit
$L, \alpha,  \beta \rightarrow \infty$ so as to regain
the known result (\ref{mag}). This has not previously been done 
directly even  for the $N=2$ Ising case: in paper I we 
calculated (\ref{mag}) from the expression for  ${\cal D}_{PQ}$ 
as the square root of an $L$ by $L$ determinant, using 
Szeg{\H o}'s  theorem.  This theorem was derived in reponse to
the first (unpublished) derivation of the Ising model 
spontaneous magnetization
by Onsager and Kaufman.\cite{Onsager1949,Onsager1971}
It was later used my Montroll, Potts and Ward.\cite{MPW1963}

Progress has been made. We have proved that the expression
III.3.45 for $S_{PQ}$ satisfies the commutation relations
 III.3.39 - III.3.41. From numerical calulations 
for small $N, L$ ($N, L  \leq 6$), it appears that these relations
(which are linear in the elements of $S_{PQ}$ , with many more
equations than unknowns) determine $S_{PQ}$ uniquely. If so,
then III.3.45 and III.3.48 have to be correct.

In paper III we defined $y_i, y'_i$ to be the
elements of the diagonal $m$ and $m'$-dimensional
matrices $Y = X_P(\alpha ) E_{PQ}$, $Y' = X_Q(\beta ) E_{QP}$.
 Both the expression III.3.48 for  ${\cal D}_{PQ}$ 
as a $2^{m+m'}\!$-dimensional sum, and the expression 
III.4.9 as an $m$-dimensional determinant, are rational functions
of the $c_i, c'_i, y_i, y'_i$. One can take all these variables to 
be arbitrary  and  can verify that the 
denominators are identical. One can  then prove that the numerators
are also the same by a recursive method using the symmetries and 
the fact that
if $c_m = c'_{m'}$, then each expression simplifies to one with
$m, m'$ replaced by $m \minus 1, m' \minus 1$.


In some ways the hardest of the three problems is to take the limit
$L, \alpha,  \beta \rightarrow \infty$. The determinant for $D_{PQ}$
is a hugely smaller calculation than the original $2^L$-dimensional 
sums in (\ref{defWPQ}), (\ref{defZP}), but it is still ultimately 
infinite. We have succeeded in calculating the determinant
in the limit $\alpha, \beta \rightarrow \infty$ as a simple product,
for finite $L, m, m'$: the key trick is to note that when 
$\alpha, \beta \rightarrow \infty$  the matrix sum in
 (\ref{defD}) can be written as the product of two Cauchy-like 
matrices.  The result (\ref{mag}) follows  by then 
taking the $L \rightarrow \infty$ limit of the product. It is this 
calculation we report here. We hope to publish the work on
the first two problems later.

\section{The matrices}
\setcounter{equation}{0}

We shall need the definitions of the $X$ and $E$ diagonal matrices. 
From  II.3.16 and II.7.4
\be \label{defX}
[X_P(\alpha)]_{i,j} \eq  \frac{-k' \sin \theta_i \sinh 
(N \alpha \lambda_i ) \; \delta_{i,j} }{\lambda_i \cosh 
(N \alpha \lambda_i) +
(1-k' \cos \theta_i ) \sinh (N \alpha \lambda_i) } \comma \ee
where
\be \label{deflam}
\lambda_i \eq (1-2 k' \cos \theta_i + k'^2) ^{1/2} \period 
\ee
The matrix $X_Q( \beta)$ is defined similarly, with $P, \alpha, 
\theta_i, \lambda_i$
replaced by $Q, \beta, \theta'_i, \lambda'_j$.

{From} II.6.18 and II.6.19, the matrix $E_{PQ}$ is an $m$ by $m$ 
diagonal matrix with entries
\be [E_{PQ} ] _{ij} \eq e( P, Q, i) \, \delta_{i,j} \comma \ee
where, for $ 0  \leq P, Q < N$ and $P \neq Q$,
 \ba  \label{defefns}
 e(P, Q, i) = &   \sin \theta_i   \;  & \! \!  {\rm if} \; \; P  < Q  
\; \; {\rm and } \; \;  m'  = m \! - \! 1  \nonumber \\
=  &  \tan ( \theta_i /2 ) \;  &   \! \!  {\rm if} \; \; P  < Q   
\;  \;  {\rm and } \;  \;  m'  =  m  \nonumber \\
=  &  1/\sin \theta_i   \; &  \! \!  {\rm if} \; \; P  > Q   \; \; 
{\rm and } \;  \;  m' =  m \! + \! 1   \\
 =  &  \cot ( \theta_i /2 ) \;   &  \! \!   {\rm if} \; \; P   > Q   
\; \;  {\rm and } \; \;  m'  =  m \period  \nonumber \ea
These equations cover all 
cases. The matrix $E (Q, P)$  is defined similarly, with 
$P , Q $ interchanged, $m, m'$ also interchanged,  and 
$\theta_i$ replaced by $\theta'_i$.

$B_{PQ}$  is an $m$ by $m'$  Cauchy-like matrix with elements
\be  \label{defB}
(B_{PQ})_{ij} \eq \frac{f_i \, f'_j}{ c_i - c'_j} \period \ee
Given $c_1, \ldots , c_m, \, c'_1, \ldots, c'_{m'}$ with 
$|m-m'| \leq 1$, there is a unique way of choosing  
$f_1, \ldots , f_m, \, f'_1, \ldots, f'_{m'}$ so that  $B_{PQ}$ 
satisfies the orthogonality condition (\ref{orth}). The working
is given in section 6 of II. We remark in section 4 of III that
it is true for arbitrary $c_i, c'_i$. Let 
\be a_i = \prod_{j=1}^{m'} (c_i - c'_j) \sep 
a'_i = \prod_{j=1}^{m} (c'_i - c_j) \comma \ee
\bd b_i = \prod_{j=1, j \neq i }^m (c_i - c_j) \sep
b'_i = \prod_{j=1, j \neq i }^{m'} (c'_i - c'_j) \comma \ed
then the results II.6.8, II.6.13,
II.6.16 can be written as
\be \label{fvals}
f_i^2 \eq \epsilon
\,a_i/b_i \; \sep 
{f'_i}^2 \eq - \epsilon
\, a'_i/b'_i \comma \ee
where $\epsilon = \pm 1$ is independent of $i$.

( For the particular values of $c_i, c'_i$ given by
(\ref{defcos}) and (\ref{defrho}), we observe numerically
that $f_i^2$ and ${f'_i}^2$ are positive real if we choose
$\epsilon = 1$ if $ P < Q$, and 
$\epsilon =  -1$ if $ P > Q$.)

A quantity that we shall need is
\be \label{defDelta}
\Delta_{m,m'} (c,c') \eq  \frac{\prod_{1 \leq i < j \leq m} 
(c_i - c_j) \prod_{1 \leq i < j \leq {m'}}  (c'_j - c'_i) }
{\prod_{i=1}^m \prod_{j=1}^{m'} (c_i - c'_j)} \period \ee


\section{The function ${\cal Z}_{P}(\alpha)$}
\setcounter{equation}{0}


The partition function $Z_P(\alpha)$ is, from II.3.16 and II. 5.38, 
or from III.3.27 and III.3.29,
\be \label{fnZpa}
Z_P(\alpha) \eq \e^{-\mu_P \alpha} \prod_{i=1}^m 
\frac{\lambda_i \cosh 
(N \alpha \lambda_i) +
(1-k' \cos \theta_i ) \sinh (N \alpha \lambda_i) }{\lambda_i}
\comma \ee
where
\be \label{defmu}
\mu_P \eq 2 k' P +(1+k') (m N- N L +L) \period \ee
When $\alpha$ is large, $Z_P(\alpha)$ has the form
$C \, \e^{g \alpha}$, where $C, g$ are independent of $\alpha$
and
\be \label{defg}
g \eq g_P \eq - \mu_P + \sum_{i=1}^m \lambda_i \period \ee
Hence from (\ref{defztilde}), ${\cal Z}_P(\alpha) 
\rightarrow C^{-1} $
as $\alpha \rightarrow \infty$.

Set 
\be \label{notn}
X_P = X_P(\infsm) \sep  x_i = (X_P)_{i,i} \; \sep
\; {\cal Z}_P  = {\cal Z}_P (\infsm)  = C^{-1} \ee
and let $c, \theta, \lambda$ be variables related to one another
by
\be \label{clam}
c \eq \cos \theta \eq (1+{k'}^2-{\lambda}^2 )/2 \, k' 
\comma \ee
then the $c_i, \theta_i, \lambda_i$ of (\ref{defcos}), 
(\ref{deflam}) are related in the same manner.
Instead of viewing $x_i$, ${\cal Z}_P$, etc. as
two-valued functions of the $c_i$, we can regard them
as single-valued rational functions of the $\lambda_i$.
Then
\be  x_i^2 \eq  \frac{{k'}^2-(1-\lambda_i)^2}
{(1+\lambda_i)^2-{k'}^2 } \comma \ee
\be \label{calZ}
{\cal Z}_P \eq \prod_{i=1}^m  \frac{4 \lambda_i}{
(1+\lambda_i)^2-{k'}^2} \eq \prod_{i=1}^m   (1+x_i^2) \period \ee
Analogous relations apply, with $P, \lambda_i, x_i$ 
replaced by  $Q, \lambda'_i, x'_i$, respectively.

Another function that we shall useful is
\be   \label{defcalR}
{\cal R} (\lambda) \eq \frac{ \prod _{i=1}^{m} 
\,(\lambda+\lambda_i)/2}
{ \prod _{j=1}^{m'} \, (\lambda+\lambda'_j)/2} \; \comma \ee
together with the elementary identity
\be \label{ident}
\left[ \frac{\Delta_{m,m'}(\lambda^2, {\lambda'}^2)}
{\Delta_{m,m'}(\lambda, \lambda')} \right]^2\eq 
2^{(m-m')^2} \prod_{i=1}^m \frac{{\cal R} (\lambda_i)}
{2 \lambda_i} \left/ \prod_{j=1}^{m'} 
{2 \lambda'_j  \, {\cal R} (\lambda'_j) } \right.\period \ee


\section{Calculation of ${\cal D}_{PQ}$}
\setcounter{equation}{0}


For any $m$ by $m'$ matrix $A$, and $m'$ by $m$ matrix $B$, it
is true that
\be \det (I_m + AB) = \det  (I_{m'} + BA) \comma \ee
so  from (\ref{defD}),
\ba \label{defD2}
{\cal D}_{PQ}(\alpha, \beta) & = &  
\det [ I_{m'} - X_Q(\beta)
E_{QP} B_{QP} X_P(\alpha) E_{PQ} B_{PQ}  ] \nonumber \\
& = &  {\cal D}_{QP}(\beta, \alpha) \period \ea
This symmetry also follows directly from the definitions 
(\ref{defS}) - (\ref{Dratio}), the fact that $\cal H$ is hermitian
and ${\cal S}_r^{\dagger} = {\cal S}_{-r} $.

 Without loss of 
generality, we can therefore 
restrict our attention to the case $P > Q$, when $m \leq m'$.

Then from (\ref{orth}) we can write $I_m$ in (\ref{defD})
as $B_{PQ} B_{PQ}^T$. Remembering that $B_{QP} = - B_{PQ}^T$, 
we can then write
(\ref{defD}) as 
\be  \label{DUB}
{\cal D}_{PQ}(\alpha, \beta) \eq \det [ U \, B_{PQ}^T ]
\comma \ee
where
\be  U \eq B_{PQ} +  X_P(\alpha) E_{PQ} B_{PQ}  X_Q(\beta)
E_{QP} \period \ee

Define 
\be  \label{defy}
 y_i =  [X_P(\alpha)]_{i,i} \, e(P,Q,i) \sep 
   y'_j =  [X_Q(\beta)]_{j,j}\, e(Q,P,j) \comma  \ee
then from (\ref{defB}) the elements of $U$ are
\be \label{defU}
U_{ij} \eq
\frac{f_i \, f'_j (1+y_i y'_j)}{ c_i - c'_j}  \period \ee

In general we do not know how to calculate the determinant 
of such a matrix. However, if we 
 take the limits $\alpha, \beta \rightarrow  
+ \infty$ and express $c_i, c'_j, y_i, y'_j$ as rational functions
of $\lambda_i, \lambda'_j$, we find that a factor 
$\lambda_i + \lambda'_j$ {\em cancels} out of the RHS of
(\ref{defU}). If $m' = m$, the result is Cauchy-like matrix, 
and one can calculate the determinant of $U$.

{\em Hereinafter we take the limit} $\alpha, \beta \rightarrow  
+ \infty$, so 
 \be \label{defy2}  y_i = x_i \, e(P,Q,i) \sep 
y'_j = x'_j \, e(Q,P,j) \comma \ee
 where $x_i, x'_j$ are given by 
 (\ref{notn}).  We write
${\cal D}_{PQ}({\infsm, \infsm}) $ simply as ${\cal D}_{PQ}$.
The integer $L$ is still finite.


\subsection*{The case $ P > Q$, $m = m'$}


The simplest case is when $m = m'$ and all matrices are square. 
so from (\ref{orth}) and (\ref{DUB}),

\be \label{ratioUB}
{\cal D}_{PQ} \eq \det U  / \det B_{PQ} \period \ee

\subsubsection*{Cauchy-like  matrices}
If $A$ is the $m$ by $m$ matrix with entries
\be \label{Cauchy}
A_{ij} \eq \frac{1}{c_i - c'_j} \comma \ee
then it is a Cauchy matrix and its determinant is
$\Delta_{m,m} (c, c')$, using the definition 
(\ref{defDelta}).\cite[eq. 2.7]{Kratten}
Any matrix with elements of the form  (\ref{defB}) is said to be 
{\em Cauchy-like}, and has determinant
\be  \label{detB}
\det B_{PQ} \eq \Delta_{m,m}(c,c') \, \prod_{i=1}^m f_i f'_i 
\ee
for all $f_i, f'_i$. We have in fact chosen the $f_i, f'_i$ so that 
$B_{PQ}$ is orthogonal, so has determinant $\pm 1$. However,  the 
form  (\ref{detB}) is convenient here as the $f_i, f'_i$ products
will cancel out of (\ref{ratioUB}).

\subsubsection*{The determinant of $U$}

From (\ref{ratioUB}), we still have to calculate the determinant of 
$U$. Its elements are given by (\ref{defU}), so $U$ is 
{\em not} in general Cauchy-like.
However, in the limit $\alpha, \beta \rightarrow \infty$ we find
that a common factor cancels from the numerator and denominator
of (\ref{defU}), and $U$ becomes Cauchy-like. We can then evaluate
its determinant by parallelling (\ref{Cauchy}) - (\ref{detB}).

The case $P > Q$, $m = m'$ is the fourth one
listed in (\ref{defefns}), so
\be (E_{PQ})_{ij} = \cot  (\theta_i/2) \, \delta_{ij} \sep 
 (E_{QP})_{ij} = \tan  (\theta'_i/2) \, \delta_{ij} \period \ee
 {From} (\ref{defX}), taking the 
limit $\alpha \rightarrow 
+  \infty$,
\be  \label{valX}
x_i \eq [X_P(\infsm)]_{ii} \eq \frac{-2 \, k' \sin \theta_i }
{(1+\lambda_i)^2-{k'}^2} \period \ee
Noting that $ \sin \theta_i \cot ( \theta_i/2) = 1+c_i = 
[(1+k')^2-{\lambda_i}^2]/2 k'$, it follows from (\ref{defy2}) that
\be y_i \eq - \, \frac{1+k'-\lambda_i}{1-k' +\lambda_i} \period \ee
(A factor $1+k' +\lambda_i$ has cancelled.)

The calculation of $y'_j$ is similar, except that now we use
$ \sin \theta'_j \tan ( \theta'_j/2) = 1 - c'_j = 
[{\lambda_j}^2 - (1-k')^2]/2 k'$ to obtain
\be y'_j \eq \frac{1-k'-\lambda'_j}{1+k' + \lambda'_j} \; \period \ee
Thus \be 1+y_i y'_j \eq  \frac{ 2 \, ( \lambda_i + \lambda'_j ) }
{(1-k' +\lambda_i) (1+k' + \lambda'_j)} \period \ee
Also, \be 
\label{clambdas}
c_i - c'_j \eq   ({\lambda'_j}^2 - {\lambda_i}^2  )/ 2 k'
\period \ee

We see that the factor $\lambda_i + \lambda'_j$ {\em cancels} out of 
(\ref{defU}), leaving
\be U_{ij} \eq  \frac{ - \, 4 \, k' \, f_i f'_j }{(1-k'+\lambda_i)
(1+k'+\lambda'_j) (\lambda_i - \lambda'_j)} \ee
so $U$ is a Cauchy-like matrix, similar to $B_{PQ}$, but
with the denominator $c_i - c'_j$ replaced by 
$\lambda_i - \lambda'_j$.  Analogously to 
(\ref{detB}), its determinant is
\be \det U \eq \Delta_{m,m}(\lambda, \lambda') \, \; \prod_{i=1}^m 
\frac{-4 k' f_i f'_i}{(1-k'+\lambda_i)
(1+k'+\lambda'_i) } \period \ee

Also, from (\ref{clambdas}), we can write (\ref{defB}) as
\be \label{formBPQ}
(B_{PQ})_{ij} \eq \frac{-2 \, k' \, f_i f'_j }
{\lambda_i^2-{\lambda'_j}^2}
\ee
so $\det B_{PQ}$ is also equal to
\be \det B_{PQ} \eq \Delta_{m,m}(\lambda^2, {\lambda'}^2 )
\prod _{i=1}^m ( -2 k' \, f_i f'_i ) \period \ee

The $f_i, f'_i$ cancel out of the ratio (\ref{ratioUB}), leaving
\be \label{Dpqres1}
{\cal D}_{PQ} \eq \frac{\Delta_{m,m}(\lambda, \lambda')}
{ \Delta_{m,m}(\lambda^2, {\lambda'}^2 )}
\;\prod_{i=1}^m \frac{2} {(1-k'+\lambda_i)
(1+k'+\lambda'_i) } \period \ee
{From}  (\ref{partfn4}), (\ref{calZ}), (\ref{defcalR})
and  (\ref{ident}) it follows that
\be \label{form1}
\left({\cal M}_r^{(2)}\right)^2 \eq
\frac{{\cal R}(1+k') \prod_{j=1}^{m'}{\cal R}
(\lambda'_j) }{ {\cal R} (1-k') \prod_{i=1}^{m}{\cal R}
(\lambda_i)  } \period \ee


\subsection*{All cases}

When $ P > Q$ and $m'  = m \! + \! 1$, we can still write
${\cal D}_{PQ}$ as in (\ref{DUB}). However, $U$ and $B_{PQ}$ are 
no longer
square matrices, so we can longer simply take products or ratios of
determinants, as in (\ref{ratioUB}). Even so, we can still
calculate ${\cal D}_{PQ}$ (and hence  ${\cal M}_r^{(2)}$) 
by adding a row to $B_{PQ}$ and  $U$ 
to make them  square Cauchy-like matrices. 
The matrix $U B_{PQ}^T$ in (\ref{DUB}) is then $m'$ by $m'$, but is 
of upper-block-triangular form. The top left block is the original 
$m$ by $m$ matrix $U B_{PQ}^T$, while the lower-right block 
is the one by one unit matrix. Hence the determinant  (\ref{DUB}) 
is unchanged and can be evaluated as a product of the two
$m'$ by $m'$ determinants.

We do this in the Appendix. The result (\ref{form2})
is  the same as (\ref{form1}), except that the factor 
${\cal R}(1+k') $ is inverted.

{From} (\ref{partfn4}) and (\ref{defD2}), ${\cal M}_r^{(2)}$
is unchanged by interchanging $P$ with $Q$, $m$ with $m'$ and 
the $\lambda_i$ with the $\lambda'_j$. {From} (\ref{defcalR}), 
this inverts the function ${\cal R}$. We can use this symmetry to 
calculate ${\cal M}_r^{(2)}$ in the other two cases.


For the four cases (\ref{defefns}), define the factor ${\cal G} $ by
 \ba  \label{defgfac}
{\cal G}  \; = &  \! \! {\cal R} (1-k') {\cal R} (1+k')   \;  & \! \! 
 {\rm if} \; \; P  < Q  
\; \; {\rm and } \; \;  m'  = m \! - \! 1  \nonumber \\
=  &   \! \! {\cal R} (1-k') /{\cal R} (1+k')  \;  &   \! \!  {\rm if}
 \; \; P  < Q   
\;  \;  {\rm and } \;  \;  m'  =  m  \nonumber \\
=  &   \! \! 1/[{\cal R} (1-k') {\cal R} (1+k')  ]  \; &  \! \!  
{\rm if} \; \; P  > Q   \; \; 
{\rm and } \;  \;  m' =  m \! + \! 1   \\
 =  &  \! \!  {\cal R} (1+k') /{\cal R} (1-k')  \;   &  \! \!   
{\rm if} \; \; P   > Q   
\; \;  {\rm and } \; \;  m'  =  m \comma  \nonumber \ea
then we find that
\be \label{allforms}
\left({\cal M}_r^{(2)}\right)^2 \eq
{\cal G}  \prod_{j=1}^{m'}{\cal R}
(\lambda'_j) \left/  \prod_{i=1}^{m}{\cal R}
(\lambda_i)   \right. \ee
for all four cases.


\section {The limit $L \rightarrow \infty$}
\setcounter{equation}{0}


The result (\ref{allforms}) is exact for finite $L$. The last step
in the calculation is to let $L \rightarrow \infty$. For this we shall 
need for the first time herein the particular definition 
(\ref{defrho}) - (\ref{defpoly})  of the polynomial (\ref{prodc}).

Let $c, \lambda$ be two variables related as are $c_i, \lambda_i$
in (\ref{clam}). Noting that $\lambda_i^2-\lambda^2 = 2 k' (c- c_i)$,
it follows from  (\ref{defcalR}) and (\ref{prodc}) that
\be  {\cal R}(\lambda) \, {\cal R}(-\lambda) \eq
(k'/2)^{m-m'} {\cal P}_P(c)/{\cal P}_Q(c) \period \ee
Using (\ref{defrho}), (\ref{defpoly}), this can be written as
\be  {\cal R}(\lambda) \, {\cal R}(-\lambda) \eq
[k'(c+1)/2]^{m-m'} z^{Q-P} \, 
{\cal W}(\lambda)  \comma \ee
where
\be  \label{Wrat}
 {\cal W}(\lambda) \eq {\cal W}_P(\lambda)/ {\cal W}_Q(\lambda)
\comma \ee
\be \label{WP}
{\cal W}_P(\lambda) \eq 1 +\sum_{n=1}^{N-1} \omega^{n(P+L)}
\left( \frac{1-z}{\omega^n-z} \right)^L  \period \ee
We have cancelled the $n=0$ term in (\ref{defrho}) from
the ratio (\ref{Wrat}).

These $z$, $w$, $c$ are related to $\lambda$ by the relations 
(\ref{defrho}), (\ref{defpoly}), (\ref{clam}). In particular,
\be \label{zwcrelns}
z^N = w = \frac{c-1}{c+1} \eq \frac{(1-k')^2- \lambda^2}
{(1+k')^2- \lambda^2} \period \ee
When $\lambda^2 < (1-k')^2$ we choose $z$ to be positive real.
Thus $w, c$ are rational functions of $\lambda$. We see that 
 $z$ is 
multi-valued, but we can choose it to be analytic by cutting 
the complex $\lambda$-plane from $1-k'$ to $1+k'$, and from
$-1-k'$ to $-1+k'$. It is then analytic on the imaginary axis 
and at infinity.

The function ${\cal W}(\lambda)$ is even and rational, its poles
and zeros being symmetrical about the imaginary axis, with none 
on the axis. Also, $z\rightarrow 1$
 and ${\cal W}(\lambda) \rightarrow 1$
as $\lambda \rightarrow \infty$, so $\log {\cal W}(\lambda)$
is analytic  in a vertical strip containing the imaginary 
axis, and tends to zero at infinity
as $1 / \lambda^{2}$. 

We can therefore perform a Wiener-Hopf 
factorization\cite{Noble1958} of ${\cal W}(\lambda)$:
\be {\cal W}(\lambda) \eq {\cal W}_{+}(\lambda) 
{\cal W}_{-}(\lambda) \comma \ee
where, for $\Re (\lambda) \geq 0$,
\be \label{logW+}
\log {\cal W}_{+}(\lambda)  \eq - \, 
\frac {1}{2 \pi \i } \int_{- \i \,\infty-
\epsilon }^{ \i \,\infty-\epsilon}
\frac{\log {\cal W}(\lambda) }{\lambda' - \lambda} \;
{\rm d}  \lambda' \ee
the integration being up a vertical line just to the 
left of the imaginary axis. This ${\cal W}_{+}(\lambda)$ is 
analytic and non-zero in the right-half 
complex $\lambda$-plane and on the imaginary axis, 
and tends to one as $\lambda \rightarrow \infty$.
The function ${\cal W}_{-}(\lambda)$ is defined
similarly, but with  $\Re (\lambda) \leq 0$ and the
sign of the RHS reversed: it is analytic and non-zero
in the LHP and on the imaginary axis.

{From} (\ref{defcalR}), the function ${\cal R}(\lambda)$
is analytic in the RHP and is proportional 
to $(\lambda/2)^{m-m'}$ when $\lambda \rightarrow \infty$.
It follows from (\ref{zwcrelns}) and
\bd k'(c+1)/2 \eq \frac{(1+k')^2-\lambda^2}{4} \comma \ed
 that
\be \label{calcR}
{\cal R} (\lambda) \eq \left( \frac{1+\lambda +k'}{2}
\right)^{m-m'}
\, \left( \frac{1-k'+\lambda}{1+k'+\lambda} \right)^{(Q-P)/N}
{\cal W}_{+}(\lambda) \period \ee
This is an exact result, true for finite $L$.


Let
\bd  t \eq \left( \frac{1-k'}{1+k'} \right)^{2/N} \comma \ed
then when $\lambda$ lies on the imaginary axis, $z$ lies on the 
positive real axis, between $t$ and 1. Hence, for $0  < n < N$,
\be \left|
 \frac{1-z}{\omega^n-z} \right| 
\leq  \left| \frac{1-t}{\omega-t } \right|   < 1\period \ee
It follows from (\ref{Wrat}), (\ref{WP}) that ${\cal W}(\lambda )$
tends uniformly to one as $L \rightarrow \infty$. For sufficiently 
small $\epsilon$, this must also be true for $\lambda$ on the 
integration line in (\ref{logW+}).
Hence 
\be {\cal W}_{+} (\lambda )  \rightarrow 1 \; \; {\rm as }\; 
\; L \rightarrow  \infty \comma \; \; {\rm for } \; 
\Re (\lambda ) \geq 0 \comma \period \ee 
so  ${\cal R}(\lambda)$ is then given by
(\ref{calcR}) with ${\cal W}_{+} (\lambda )  = 1$.

It follows that
\be \prod_{j=1}^{m'}{\cal R}
(\lambda'_j) \left/  \prod_{i=1}^{m}{\cal R}
(\lambda_i)   \right. \eq {\cal R} (1+k')^{m'-m} \, \left[ 
\frac{{\cal R} (1+k')}
{{\cal R} (1-k') }\right]^{(Q-P)/N} \comma \ee
\be {\cal R}(1+k') \eq (1+k')^{m-m'+(P-Q)/N}
\sep
{\cal R}(1-k') \eq (1-k')^{(Q-P)/N} \period \ee

Using $r = Q-P$, modulo $N$, we obtain
\be \label{final}
\left({\cal M}_r^{(2)}\right)^2 \eq (1-k'^2)^{r (N-r)/N^2} \ee
for all four cases, in agreement with (\ref{mag}).

\subsection*{Asymptotic degeneracy}

We are now in a position to confirm the remarks we made before 
(\ref{partfn3}). Expanding (\ref{defcalR}) and (\ref{calcR}) to
first order in $1/\lambda$ for $\lambda$ large and equating
the coefficients, using (\ref{defmu}) and (\ref{defg}),
we obtain
\bd \label{sumlam}
g_P \eq g_Q \period \ed

We noted after (\ref{defg}) that when $\alpha$ is large, then
$Z_P(\alpha)$ is proportional to $\exp( g_P \alpha)$, so 
$Z_P(\alpha)/Z_Q(\alpha)$ must tend to a limit 
as $\alpha \rightarrow \infty$. From(\ref{defztilde}) and (\ref{notn}),
this is ${\cal Z}_Q/{\cal Z}_P$, so from (\ref{calZ}) and (\ref{defcalR}),
\be 
\lim_{\alpha, L \rightarrow \infty} \frac{Z_P(\alpha)}{Z_Q(\alpha)} 
\eq \frac{{\cal R}(1-k') {\cal R}(1+k')}
{2^{m-m'} \, {\cal R}(0) } \eq 1  \comma\ee
using (\ref{calcR}) with ${\cal W}_{+}(\lambda) = 1$ to evaluate the 
middle expression.


Hence in the limit $\alpha,\beta,  L \rightarrow \infty$, $Z_P(\alpha)$
is the same for all $P$. We have already shown in (\ref{final})
that ${\cal M}_r^{(2)}$ depends on $P$ and $Q$ only via the difference
$r$. {From} (\ref{Dratio}) - (\ref{partfn4}), the same must be true
of $W_{PQ}(\alpha,\beta)$. This justifies our replacing the sums in 
(1.12) (for $b=0$) by single terms.


\section {Summary}
\setcounter{equation}{0}

The magnetization ${\cal M}_r$ of the chiral Potts 
model can be expressed as the expectation value of $\omega^{r a}$,
where $a$ is a spin inside a cylindrical lattice with the spins at 
the top and bottom boundaries fixed to zero. This was calculated 
in \cite{RJB2005a,RJB2005b} analytically by generalizing  ${\cal M}_r$,
showing that it satisfied  functional relations in the large-lattice 
limit, and then solving those relations.
 This was quite a different method from the algebraic techniques
 originally used for 
the Ising model. 

To obtain ${\cal M}_r$ for the general solvable 
model it is sufficient to obtain it for the superintegrable case, and
this case has many resemblances to the Ising model. This naturally leads
to the question whether there is an algebraic way of calculating the 
magnetization for the superintegrable chiral Potts model.

We looked at this problem in three previous 
papers,\cite{paper1}-\cite{paper3} which we refer to as I, II, III.
In I we revisited the Ising model (the $N=2$ case of the 
superintegrable model) and showed that ${\cal M}_r$ was proportional 
to the determinant I.7.7 and I.7.9.


In II  we first showed in II.5.37 that ${\cal M}_r$ was proportional 
to the weighted sum   of the elements of a  $2^m$ by $2^{m'}$ matrix 
${\cal S}_{\rm red}^r$.
Generalizing our calculation in I, we then
conjectured that it was proportional to the $m$ by $m'$ determinant
${\cal D}_{PQ} (\alpha,\beta)$ of (\ref{defD}) herein.

In III we further showed that the matrix 
$S_{PQ} = {\cal S}_{\rm red}^r$ satisfied a  number of 
commutation relations, in particular the two relations (2.21), 
(2.22) therein, and 
further conjectured that the elements of $S_{PQ}$ had
the particular simple product form III.3.45.

We have since proved that this form for $S_{PQ}$
both satisfies the two commutation relations and implies
the determinantal result. We hope to publish the working
soon.  What we have not done, and would be  needed to complete 
the proof, is to show that the commutation relations III.3.39 
and III.3.40 (plus the simple normalization property III.3.41) 
define $S_{PQ}$ uniquely. Calculations
for small lattices suggest that this is so.

These calculations are all for finite sums and determinants of 
finite matrices. They involve parameters $\alpha, \beta$, which 
can be regarded as a measure of the number of rows below and 
above the selected spin $a$ in the lattice. They also involve the 
lattice width $L$. Ultimately we want to take the limits
$\alpha, \beta, L \rightarrow \infty$.


To complete this algebraic calculation of ${\cal M}_r$, we need 
to evaluate the determinant ${\cal D}_{PQ} (\alpha,\beta)$.
We do not know how to do this for finite $\alpha, \beta$, but we show 
in section 4 and the Appendix herein that in the limit when 
$\alpha$ and $\beta$ are both infinite,  
${\cal D}_{PQ} (\alpha,\beta)$ can be written as 
the determinant of the product of two square Cauchy-like matrices. 
We can therefore evaluate the determinant as a product of terms, 
the number of terms being quadratic in $m, m'$.
Finally in section 5 we take the limit $L \rightarrow \infty$, 
showing that the needed function ${\cal R}(\lambda)$ then has a 
simple form for $\Re (\lambda) \geq 0$. We of course regain the 
result (\ref{mag}) of  \cite{RJB2005a,RJB2005b},
which had been conjectured by Albertini 
{\it et al} in 1989.\cite{AMPT89}

For the $N \! = \! 2$ Ising case, this appears to be a new way of 
calculating the needed determinant. Yang\cite{Yang1952} did so by 
calculating the  eigenvalues in the large-$L$ limit. 
Montroll {\it et al},\cite{MPW1963} and  presumably Onsager and 
Kaufman in  1949,\cite{Onsager1949, Onsager1971} did so by expressing
the result in terms of a Toeplitz determinant and then using 
Szeg{\H o}'s theorem.\cite{Szego1958}


\section {Additions}
\setcounter{equation}{0}
Since posting this paper on the Los Alamos archive, there has been 
considerable further progress on the first two problems listed above 
in the  Introduction. Iorgov {\it et al}\cite{Iorgov2009}
 have proved that $S_{PQ}$ is 
indeed  given by III.4.9, and therefore $D_{PQ}$ by III.3.48. They 
did this by showing that III.3.45 satisfied the commutation relations
III.3.39 - III.3.41, and were able to prove that these relations
have a unique solution. Further, they went on  to calculate 
$D_{PQ}$ in the limit $\alpha, \beta \rightarrow \infty$ directly,
using the expression III.3.48 as a sum over matrix elements,
rather than the determinantal form III.4.9 or III.4.10.

The author has also posted a paper on the archive\cite{paper5}
 giving the proof of 
the equivalence of the sum and determinantal forms of $D_{PQ}$ that 
is outlined in the penultimate paragraph of the introduction above.
It is remarked therein that the motivation for this work
was the similarity between the Ising and superintegrable chiral 
Potts models, in particular to find a derivation of the determinant
form of $D_{PQ}$. This determinantal form  was conjectured in II 
by generalizing the Ising model result as expressed in I.


\section {Acknowledgements}
\setcounter{equation}{0}


The author thanks Helen Au-Yang and Jacques Perk for many helpful 
discussions during their stay at the Australian National University
in 2008/2009.  He also thanks the Rockefeller Foundation for a residency 
at Bellagio, Italy, that inspired this work. It is partially 
supported  by the Australian  Research Council.



\section*{Appendix  A}
\renewcommand{\theequation}{A\arabic{equation}}
\appendix
\setcounter{equation}{0}

Here we consider the case when $P > Q$ and $m' = m+1$, so the matrices
$B_{PQ}$, $U$ are not square. Then from (\ref{defefns}),
$e(P,Q,i) = 1/\sin \theta_i$ and $e(Q,P,j) = \sin \theta'_j$. Using 
(\ref{defy}), (\ref{valX}), we obtain
\be y_i  = \frac{-2 k'}{(1+\lambda_i)^2-{k'}^2}
\, \sep y'_j  = \frac{(1-\lambda'_j)^2-{k'}^2}{2 k'} \period \ee
Hence \be 1+ y_i y'_j \eq \frac{(\lambda_i +\lambda'_j)
(2+\lambda_i -\lambda'_j )}{(1+\lambda_i)^2 -{k'}^2} \period \ee
Using (\ref{clambdas}), we see that
the factor $\lambda_i +\lambda'_j$ again cancels in (\ref{defU}),
leaving
\be \label{usum}
U  \eq  U^{(1)}  +  U^{(2)}   \comma \ee
where $U^{(1)}, U^{(2)}$ are matrices with elements
\be \label{defU1}
U_{ij}^{(1)} \eq \frac{-4 k' f_i f'_j }
{(\lambda_i -\lambda'_j) [(1+\lambda_i)^2-{k'}^2)]} \comma \ee
\be U_{ij}^{(2)} \eq \frac{-2 k' f_i f'_j }
{ (1+\lambda_i)^2-{k'}^2)} \comma \ee
respectively. Thus
\be  U^{(2)} \eq  {\xi} \, {\eta}^T \comma \ee
where ${\xi}, \eta$ are vectors and  $({\eta})_j = f'_j$.

Consider the vector
$  B_{PQ} \, {\eta} $:
from (\ref{defB}) it has entries
\bd   (B_{PQ} \, {\eta})_i \eq   f_i \, {\cal F}'(c_i) \comma \ed
where
\bd {\cal F}'(c)  \eq \sum_{j=1}^{m'}  \frac{{f'_j}^2}{c - c'_j} 
\period \ed
Remembering that $P > Q$, this is precisely the sum considered in
 (II.6.6), but with $p, q$ therein replaced by $Q, P$, so 
(also using II.6.14),
\be \label{sumfp}
{\cal F}'(c)  \eq  \gamma' + 
  {\cal P}_P(c)/{\cal P}_Q(c) \comma      \ee 
where $\gamma'$ is independent of $c$.
Taking the limit $c \rightarrow \infty$, we obtain $\gamma' = 0$.
It follows that ${\cal F}'(c) $ 
vanishes when $c = c_i$ ($ i = 1, \ldots, m$), so 
\be \label{borthf}
B_{PQ} \, {\eta}  \eq 0 \period \ee
Substituting (\ref{usum}) into (\ref{DUB}), the $U^{(2)}$ term
is zero, leaving
\be \label{DUB1}
{\cal D}_{PQ} \eq \det [ U^{(1)} B_{PQ}^T ] \period \ee

We have reduced the problem to one of calculating the determinant
of a product of two Cauchy-like matrices, but unfortunately
they are not square. The solution to this problem is actually 
suggested by (\ref{borthf}): of we define an $m'$ by $m'$  matrix
\be {\cal B} \eq \left(\begin{array} {c} B_{PQ} \\ {\eta}^T
\end{array} \right)  \comma \ee
(dropping the suffixes $P,Q$).
Then, using (\ref{orth}), all its $m'$ rows are mutually orthogonal.
Multiplying (\ref{sumfp}) by $c$ and taking the limit 
$c \rightarrow \infty$, we obtain
\be \label{ff}   {\eta}^T \eta \eq \sum_{j=1}^{m'} {f'_j}^2 
\eq   1 \comma \ee
so together with (\ref{borthf}) it follows that
${\cal B} {\cal B}^T = I_{m'}$, i.e. $\cal B$ is a square 
orthogonal matrix.

We also extend $U^{(1)}$ by adding the row $\eta^T$ to form
the square matrix
\be {\cal U} \eq \left(\begin{array} {c} U^{(1)} \\ \eta^T
\end{array} \right)  \period \ee
Then, using (\ref{borthf}) and (\ref{ff}), 
\be {\cal U}  \, {\cal B}^T \eq \left(\begin{array} 
{cc} U^{(1)} B_{PQ}^T  & U^{(1)} \eta\\
\eta^T B_{PQ}^T & \eta^T \eta
\end{array} \right) 
\eq \left(\begin{array} {cc} U^{(1)} B_{PQ}^T  & U^{(1)} \eta \\
 {\bf 0} &  1
\end{array} \right)  \period \ee
We see that ${\cal U \cal B}^T$ is an upper-right block triangular 
matrix and
 $\det {\cal U}  \, {\cal B}^T  = \det U^{(1)} B_{PQ}^T$. {From}
(\ref{DUB1}), using the orthogonality of $\cal B$,
\be \label{Drat}
{\cal D}_{PQ} \eq  \det {\cal U} /\det {\cal B} \period \ee

The square matrices $\cal B$ and $\cal U$ are Cauchy-like.
All the elements ${\cal B}_{ij} $  of $\cal B$ are given 
by the RHS of (\ref{formBPQ}), provided
 we take $f_{m+1} = -\lambda_{m+1}^2/2 k'$ and then let 
$\lambda_{m+1} \rightarrow \infty$.  Using the general formula 
(\ref{detB}) for an $m'$ by $m'$ determinant, and then
taking this limit, we obtain
\be \det {\cal B} \eq \Delta_{m,m'} (\lambda^2,{\lambda'}^2) \, 
\prod_{i=1}^m (2 k' f_i) \, \prod_{j=1}^{m'}  f'_j \comma \ee
where  $\Delta_{m,m'} (c,c') $ is defined by (\ref{defDelta}).
(Similarly to section4, we know from the orthogonality 
of $\cal B$ that its determinant in $\pm 1$, but this form is 
convenient here because the $f_i, f'_j$ products will cancel out of
(\ref{Drat}).)


Similarly, all the elements ${\cal U}_{ij} $ of ${\cal U}$ are given 
by the RHS of (\ref{defU1}), except that now we take
$f_{m+1}  =  -\lambda_{m+1}^3/4 k'$ before letting
$\lambda_{m+1} \rightarrow \infty$. Again using the general formula 
(\ref{detB}), we find that
\be \det {\cal U} \eq \Delta_{m,m'} (\lambda,\lambda') \, 
\prod_{i=1}^m \frac{4 k' f_i}{(1+\lambda_i)^2-{k'}^2}  
\, \prod_{j=1}^{m'}  f'_j \period  \ee

Hence from (\ref{Drat}),
\be {\cal D}_{PQ} \eq \frac{\Delta_{m,m'} (\lambda, \lambda')}
{\Delta_{m,m'} (\lambda^2, {\lambda'}^2)} \;
\prod_{i=1}^m \frac{2} {(1+\lambda_i)^2 - {k'}^2}
\period \ee
Using (\ref{partfn4}), (\ref{calZ}), (\ref{defcalR})
and  (\ref{ident}), we now obtain
\be \label{form2}
\left({\cal M}_r^{(2)}\right)^2 \eq
\frac{\prod_{j=1}^{m'}{\cal R}
(\lambda'_j) }{ {\cal R} (1+k')\, {\cal R} (1-k') \, \prod_{i=1}^{m}
{\cal R} (\lambda_i)  } \period \ee
This is the result quoted at the end of section 4.



 \end{document}